\documentclass[preprint]{article}
\usepackage{graphicx,amsmath,bm}
\usepackage{color}
\usepackage[normalem]{ulem}
\begin{document}
%\emergencystretch 3em
%%paper title
\title{Models for membrane curvature sensing of curvature generating proteins}

\author{T. V. Sachin Krishnan\textsuperscript{1,\footnote{sachin@physics.iitm.ac.in}}, Sovan L. Das\textsuperscript{2}\and P. B. Sunil Kumar\textsuperscript{1,2}}
\date{{\small \textsuperscript{1} Department of Physics, Indian Institute of Technology Madras, Chennai, India\\
\textsuperscript{2} Department of Mechanical Engineering, Indian Institute of Technology Palakkad, Kerala, India\\
\textsuperscript{3} Department of Physics, Indian Institute of Technology Palakkad, Kerala, India}\\
  \today}% It is always \today, today,

%%author names are separated by comma (,)
%%use \and before the last author name
%%\textsuperscript{number} is used for affiliation
%%use a * along with the number separated by comma
%% for the  author for correspondence

%%escape two column mode for title, affiliation and abstract
%%by giving \twocolumn command as shown

%\twocolumn[{%

\maketitle

%%include \corres to print the corresponding author Email id
%\corres{sachin@physics.iitm.ac.in}

%%include \msinfo for
%%manuscript information such as
%%received, revised and accepted dates
%%
%\msinfo{15 July 2019}{}{}

%%abstract
\begin{abstract}
  The curvature sensitive localization of proteins on membranes is vital for many cell biological processes.
   Coarse-grained models are routinely employed to study the curvature sensing phenomena and membrane morphology at the length scale of few micrometers.
   Two prevalent phenomenological models exist for modeling experimental observations of curvature sensing, (1) the spontaneous curvature model and (2) the curvature mismatch model, which differ in their treatment of the  change in elastic energy due to the binding of proteins on the membrane.
   In this work, the prediction of sensing and generation behaviour, by these two models,  are investigated using analytical calculations as well as Dynamic Triangulation Monte Carlo simulations of quasi-spherical vesicles.
  While the spontaneous curvature model yields a monotonically decreasing sensing curve as a function of vesicle radius, the curvature mismatch model results in a non-monotonic sensing curve.
  We highlight the main differences in the interpretation of the protein-related parameters in the two models.
We further propose that the spontaneous curvature model is appropriate for modeling peripheral proteins employing the hydrophobic insertion mechanism,  with minimal modification of membrane rigidity, while the curvature mismatch model is appropriate for modeling curvature generation using scaffolding mechanism where there is significant stiffening of the membrane due to protein binding.

\end{abstract}

%%insert keywords separated by comma using \keywords{words}
%\keywords{biological membranes, curvature sensing, curvature generation}

%%include \pacs{number} to print the PACS number
%\pacs{}

%}]
%%close the twocolumn escape here

%%include \doinum{number}for the DOI number in the header
%%include \volnum{number} for the volume number in the header
%%include \year{yyyy} for  year of publication in the header
%%include \pgrange{num--num} page range of article in the header
%%include \artcitid{num} for the article citation id
%%include \lp to print last page of the article
%%include \setcounter{page}{pagenum} for the exact starting page of the article

%\doinum{}%12.3456/s78910-011-012-3}
%\artcitid{\#\#\#\#}
%\volnum{1}
%\year{2019}
%\pgrange{1--7}
%\setcounter{page}{1}
%\lp{7}

%%%%%%%%%%%%%%%%%%%%%%%%%%%%%%%%%%%%%%%%%%%%%%%%%%%%%%%%%%%%%%%%%%%%%%%%%%%%%%
\section{Introduction}
%%%%%%%%%%%%%%%%%%%%%%%%%%%%%%%%%%%%%%%%%%%%%%%%%%%%%%%%%%%%%%%%%%%%%%%%%%%%%%
\label{sec:introduction}
Protein mediated regulation of membrane curvature occurs during many cellular processes such as cargo trafficking, cell motility, cell growth, and division \cite{McMahon2005, Zimmerberg2006, Jarsch2016, Bassereau2018}.
Recently, several classes of proteins capable of curvature generation have been identified~\cite{Farsad2003, Voeltz2007, Shibata2009}.
Dynamin and proteins with the crescent-shaped Bin Amphiphysin Rvs (BAR) domain were found to generate curvature by the scaffolding mechanism \cite{Peter2004}.
On the other hand, epsin protein with an N-terminal helix generates curvature using the hydrophobic insertion mechanism~\cite{Bhatia2009, Hatzakis2009}.
These curvature generating proteins are also capable of sensing membrane curvature~\cite{Antonny2011}.
Curvature sensing refers to the ability of proteins to bind onto membranes depending on the local curvature.
Recent experiments have reported that membrane curvature gives a cue to localization of proteins in bacteria and viruses~\cite{Wasnik2015, Gill2015, Martyna2016, Draper2017}.
This phenomenon is believed to be exploited by cells during the process of  budding and fission.
For example, in the clathrin-mediated membrane fission, the narrow neck between the clathrin-bound bud and the parent membrane preferentially recruits the dynamin proteins responsible for membrane scission.
Thus, it is important to understand  these processes of curvature sensing and generation to gain insight into  many of the cellular processes.

\begin{figure}
  \centering
  \includegraphics[width=0.6\textwidth]{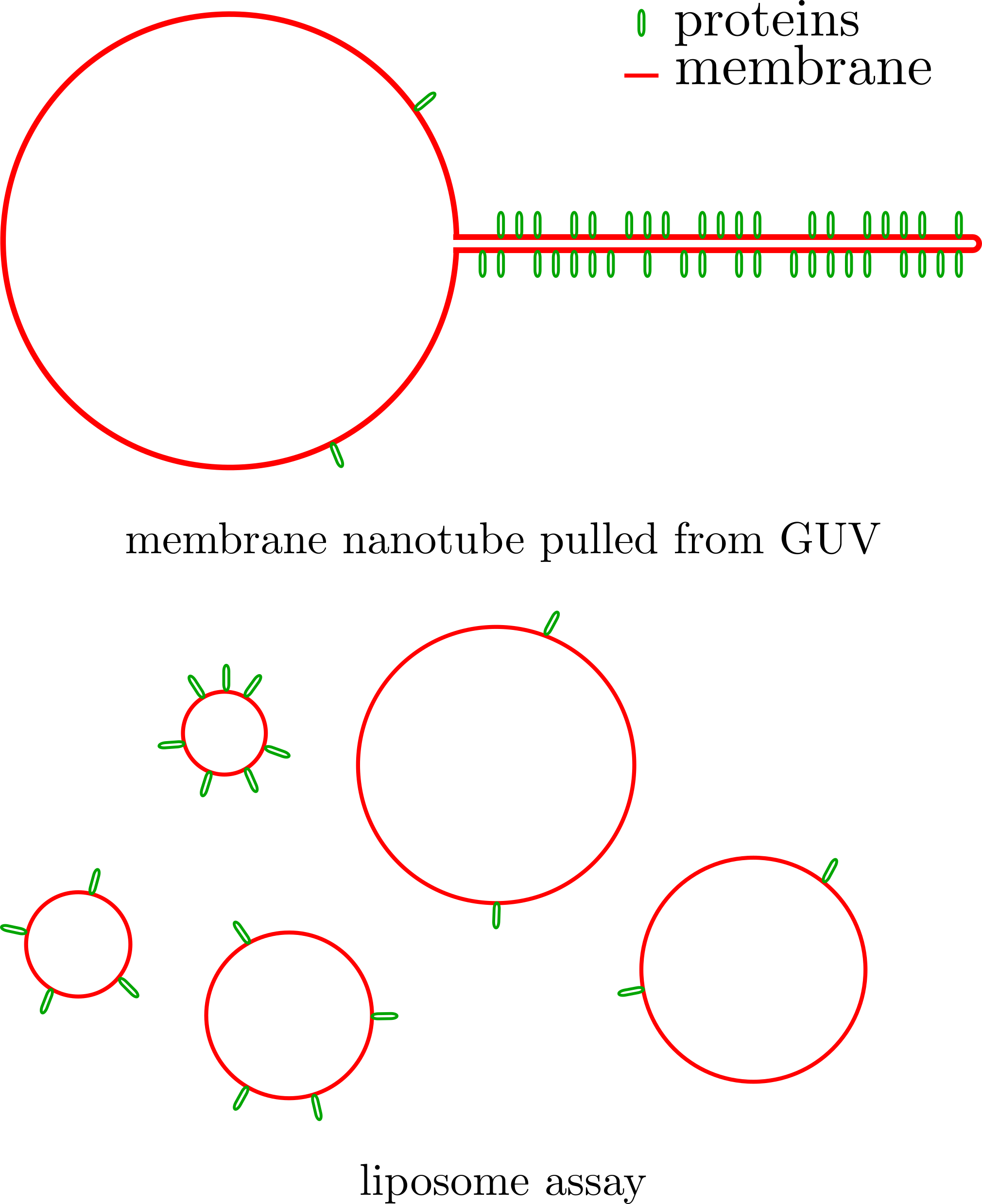}
  \caption{Preferential binding of proteins to highly curved membrane surfaces.
    Schematic of typical biophysical experimental setups used to study curvature sensing phenomena.
    }\label{fig:schematic}
\end{figure}

Biophysical experimental setups such as Single Liposome Curvature (SLiC) assays and tethers pulled from giant unilamellar vesicles (GUVs) have been extensively used to quantify curvature sensing~\cite{Baumgart2011, Aimon2014}.
These two methods are schematically illustrated in Fig.~\ref{fig:schematic}.
Considering their small throughput, \textit{in vivo} alternatives have also been used~\cite{Rosholm2017}.
In the tether pulling experiments, a narrow membrane tube of few tens of  nanometer radius is pulled from a GUV of a few microns radius.
Curvature sensitive proteins are introduced to these two membrane surfaces with very different curvatures.
The relative binding fraction of proteins on the two surfaces is then measured based on the intensity of fluorescently tagged proteins.
On the other hand, in SLiC assays, proteins are introduced in a medium containing liposomes of different radii~\cite{Bhatia2009}.
Similar to the case of tether pulling experiments, the intensity of fluorescently tagged proteins is utilized to estimate the binding fraction of proteins on liposome surfaces.

Several quantitative analytical models have also been proposed to study the phenomena of curvature sensing~\cite{Zhu2012, Bozic2015, Svetina2015}.
Currently, there exist two thermodynamic models for curvature sensing/generation --- the spontaneous curvature model and the curvature mismatch model.
The two models differ in their treatment of the membrane elastic energy due to the deformation induced by proteins.
Although both the models have been shown to fit various experimental data, it is not clear which among the two models is most suitable to study the curvature sensing/generation behavior of a particular  protein.
In the present  work, we describe the two models and compare the results obtained using analytical calculations as well as monte carlo simulations.

The article is organized as follows.
Section~\ref{sec:models} introduces the two thermodynamic models and presents analytical results.
In sec.~\ref{sec:simulations}, we discuss the sensing/generation behaviour of the two models studied using MC simulations.
The article ends with a few concluding remarks  in sec.~\ref{sec:conclusions}

\section{Models for curvature sensing}
\label{sec:models}
The curvature sensing ability of proteins is a consequence of the interaction between the proteins and the membrane.
Although the specific interactions between membrane patches and protein domains are quite complicated, their effects can be understood in terms of a few coarse-grained interaction parameters.
At mesoscopic length scales, several quantitative analytical models have been proposed to study the phenomena of curvature sensing.
Below we describe and compare two  models, which are most often used.

\subsection{Spontaneous Curvature Model}
\label{sec:spcur_model}
Spontaneous Curvature (SC) model assumes that the only effect  of the bound protein, on the elastic energy,    is to  induce a  preferred  local  curvature of  the membrane.
This model has been employed previously to study sorting of amphiphysin in tube pulling assays~\cite{Sorre2012} as well as in modeling of lipopolysaccharide-binding on synthetic lipid vesicles~\cite{Mally2017}.
In this model, the energy of the membrane surface is given by the spontaneous curvature form of the Helfrich free energy~\cite{Helfrich1973},
\begin{align}
  \mathcal{H} = \int \textrm{d}A \frac{\kappa}{2} \left(2H - C_0 \right)^2,
  \label{eq:ci_energy}
\end{align}
where $\kappa$ is the bending rigidity and $C_0$ is the membrane spontaneous curvature.
The integral is over the entire area of the membrane surface.
The spontaneous curvature is usually assumed to be linearly dependent on the protein area fraction $\phi$~\cite{Markin1981, Leibler1986},
\begin{align}
  C_0 = C_p \phi,
\end{align}
where $C_p$ is the intrinsic curvature of the protein.
In essence, this model assumes that the protein sets a preferred local curvature on the membrane depending on its bound density.
\begin{figure*}
  \centering
  \includegraphics[width=\textwidth]{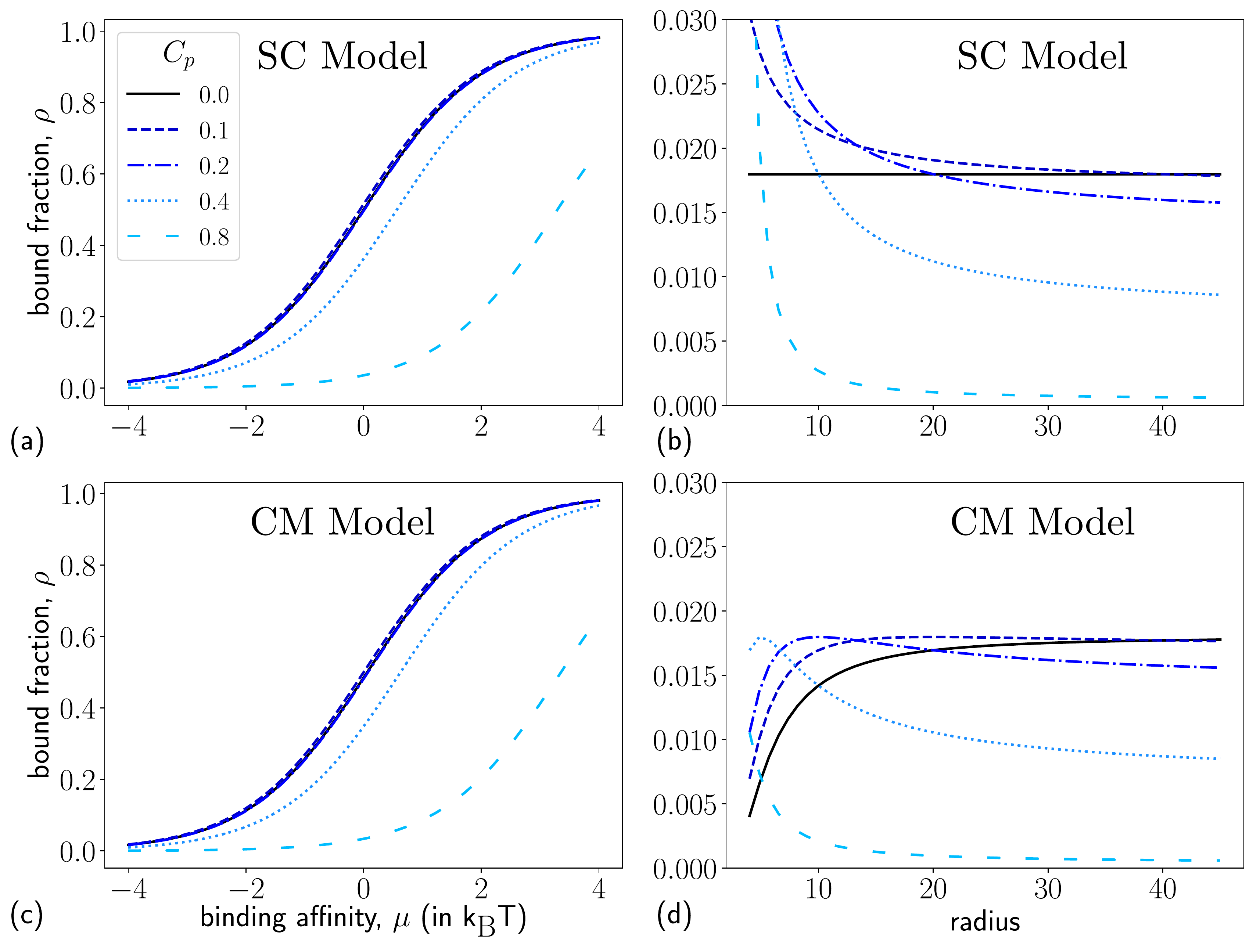}
  \caption{Protein binding on a non-deformable spherical vesicle studied  using the spontaneous curvature model and the curvature mismatch model.
    Adsorption isotherms of proteins with different $C_p$ on vesicle of size $R = 21$ in (a) the SC model and (c) the CM model.
    Curvature sensing curves for proteins of various $C_p$ values at $\mu = -4$ for (b) the SC model and (d) the CM model.}\label{fig:analytical}
\end{figure*}

We consider the vesicle as a triangulated surface with $N_v$ vertices.
A discretized Hamiltonian for this surface can be written as,
\begin{align}
  \mathcal{H}_{\textrm{SC}} &= \frac{\kappa}{2} \sum_{i=1}^{N_{v}} (2H_i - C_p\phi_{i})^2 A_{i} - \mu \sum_{i=1}^{N_{v}} \phi_{i},
  \label{eq:sc_hamiltonian}
\end{align}
where $H_{i}$ and $\phi_{i}$ are respectively the mean curvature and the protein-bound state at vertex $i$.
The concentration of proteins in bulk is taken into account indirectly through the binding affinity parameter $\mu$.
The parameter $\mu$ is the free energy of the proteins in the reservoir for binding onto the membrane surface.
  It depends on the interaction energy between the membrane and the protein
  and also the concentration of the protein in the bulk ($c_{\textrm{bulk}}$) through the relation,
\begin{align}
  \mu = \mu_0 + \log \frac{c_{\textrm{bulk}}}{c_0}
\end{align}
where $\mu_0$ and $c_0$ are, respectively, the standard state protein chemical potential and concentration~\cite{Sachin2019}.

For small bound fractions, the proteins do not significantly affect the membrane curvature if they are homogeneously distributed over the surface.
Therefore, we can simplify the expression for free energy by assuming a perfectly spherical surface with each vertex having the same curvature ($2H$).
For such  a uniformly spherical surface, the mean curvature at each vertex is simply the inverse of the vesicle radius, \textit{ie.} $H_{i} = H = 1/R$.
The variable $\phi_i$ takes value one in vertices with a bound protein and zero in others.
In this model, protein-bound vertices will have minimum energy when the local curvature matches with protein intrinsic curvature.
Further, if the area at each vertex $A_{i}$ is same, say $a$, we can write an effective free energy per-vertex as a function of the protein bound fraction $\rho = N_{p} / N_{v}$ as
\begin{align}
  f_{\textrm{SC}} (\rho) &= \frac{\kappa a}{2} \left[ \left(2H\right)^2 (1 - \rho) + \left(2H - C_p\right)^2 \rho \right] - \mu \rho \nonumber \\
  &+ k_{\textrm B}T \left[ \rho \log (\rho) + (1 - \rho) \log (1 - \rho) \right], 
  \label{eq:effective_ci_energy}
\end{align}
where, $N_p=\sum_{i=1}^{N_v} \phi_i$, is the total number of vertices occupied by the protein field.
The first term is obtained as a result of separating the sums for vertices with and without proteins in Eq.~(\ref{eq:sc_hamiltonian}).
The last term in Eq.~(\ref{eq:effective_ci_energy}) represents the mixing free energy of proteins on the discretized surface.
Note that such a mixing free energy is due to the exclusion interaction of proteins on the discretized surface.
The protein bound fraction in equilibrium is obtained by minimizing the effective free energy wrt $\rho$ as
\begin{align}
  \rho_{\textrm{eq}} = \frac{1}{1 + e^{-\beta\left[ \mu - \frac{\kappa a C_{p}}{2} \left(C_p - 4H \right) \right]}}.
\label{eq:shifted_langm}
\end{align}
When $C_p = 0$, the above equation takes the form of standard Langmuir isotherm.
For non-zero $C_p$ values, the Langmuir isotherm is recovered by defining an effective chemical potential,
\begin{align}
\mu' = \mu - \frac{\kappa a C_p}{2} \left( C_p - 4H \right).
\end{align}
The adsorption isotherms for different $C_p$ for the  SC model is shown in Fig.~\ref{fig:analytical}a.
The  isotherms for non-zero spontaneous curvatures are shifted Langmuir isotherms as predicted by Eq.~(\ref{eq:shifted_langm}).
Experiments have reported that the  adsorption  of some proteins on vesicles follows the Langmuir isotherm~\cite{Bhatia2009}.
The preferential binding of proteins to vesicles of various sizes is characterized using a curvature sensing curve,
wherein the bound fraction of protein  is plotted against the vesicle size at a particular binding affinity.
The curvature sensing curve at $\mu = -4$  is shown in Fig.~\ref{fig:analytical}b.
When $C_p = 0$, the protein bound fraction does not depend on the vesicle radius as there is no coupling between the mean curvature $H$ and the protein bound fraction $\phi$ in Eq.~(\ref{eq:ci_energy}).
Therefore, within the SC model, $C_p = 0$ corresponds to the case where membrane curvature  is  insensitive to that of protein.
For non-zero $C_p$, the bound fraction increases with decreasing vesicle radius; approaching the maximum of $1$ as  $R \rightarrow 0$ (or $H \rightarrow \infty$ in Eq.~(\ref{eq:shifted_langm})).
Note that, in the SC model, the protein bound fraction monotonically decreases with increasing vesicle size.

\subsection{Curvature Mismatch Model}
\label{sec:mismatch_model}
Curvature mismatch (CM) model supposes a) an energy penalty when there is a difference in the local membrane and protein curvatures and b)  curvature stiffness of the membrane to depend on the local protein concentration.
It has successfully reproduced the preferential binding of I-BAR proteins to negatively curved membranes~\cite{Prevost2015}, sorting of potassium channel KvAP~\cite{Bozic2015}, and sorting of transmembrane proteins in live cell filopodia~\cite{Rosholm2017}.
In the curvature mismatch (CM) model, the Hamiltonian is of the form,
\begin{align}
  \mathcal{H} = \int \textrm{d} A \left[ \frac{\kappa}{2} \left(2H\right)^2 + \frac{\bar{\kappa}}{2} \left(2H - C_p \right)^2 \phi \right].
  \label{eq:cm_energy}
\end{align}
Here the first term is the Helfrich energy for the membrane surface and the second term is the curvature mismatch energy.
The parameter $\bar{\kappa}$ decides the strength of the mismatch penalty.
In regions where there are no bound proteins, $\phi = 0$, only the first term in Eq.~(\ref{eq:cm_energy}) contributes to the energy.
In this limit of no bound protein, both spontaneous curvature model and the curvature mismatch model have the same Hamiltonian.

In order to compare the CM model with the SC model discussed previously, we derive the equilibrium protein bound fraction on a non-deformable vesicle.
The discretized form of the CM free energy model is given by,
\begin{align}
  \mathcal{H}_{\textrm{CM}} &= \sum_{i=1}^{N_{v}} \left[ \frac{\kappa}{2} (2H_i)^2 + \frac{\bar{\kappa}}{2} (2H_i - C_p)^2 \phi_i \right] A_{i} - \mu \sum_{i=1}^{N_{v}} \phi_{i}.
  \label{eq:cm_hamiltonian}
\end{align}
In this model, protein bound vertices will have the same energy as that for the unbound vertices, when the local curvature matches with the protein curvature.
As in Eq.~(\ref{eq:effective_ci_energy}), the effective free energy for the CM model in terms of the bound fraction $\rho$ is
\begin{align}
f_{\textrm{CM}} (\rho) &= \frac{\kappa a}{2} (2H)^2  + \frac{\bar{\kappa}a}{2} (2H - C_p)^2 \rho - \mu \rho \nonumber \\
&+ k_{\textrm B} T \left[ \rho \log (\rho) + (1 - \rho) \log (1 - \rho) \right].
\end{align}
For simplicity, we assume that $\kappa = \bar{\kappa}$ in the rest of the discussion.
The equilibrium bound fraction obtained after minimizing the effective free energy with respect to  $\rho$ is,
\begin{align}
  \rho_{\textrm{eq}} = \frac{1}{1 + e^{-\beta\left[ \mu - \frac{\kappa a}{2} \left(2H - C_p \right)^2 \right]}}.
\end{align}
Here again, the binding assumes the form of a shifted Langmuir isotherm.
However, the effective binding affinity is different from that obtained for the SC model.
For the CM model, the effective binding affinity takes the form,
\begin{align}
\mu' = \mu - \frac{\kappa a}{2} \left( 2H - C_p \right)^2.
\end{align}
Here, the effective binding affinity is quadratic in the vesicle curvature, with an additional term $-2 \kappa H^2 a$.
This is unlike in the SC model, where the dependence on curvature is linear.

For large vesicle radius (small $H$),  we see no difference in the adsorption isotherm obtained with the two models (see Fig.~\ref{fig:analytical}a and Fig.~\ref{fig:analytical}c).
The additional quadratic term in the effective binding affinity of CM model becomes relevant when the vesicle size is small (large $H$).
Consequently, the curvature sensing curves predicted using the two models differ significantly as seen in Fig.~\ref{fig:analytical}b and Fig.~\ref{fig:analytical}d.
While the SC model predicts a monotonic inverse relation between the protein bound fraction and the vesicle radius, the CM model predicts a non-monotonic dependence.
Since, the additional term is negligible at small vesicle curvatures ($H \ll C_p$), the predictions from the two models are similar for larger vesicles.

\begin{figure*}
  \centering
  \includegraphics[width=\textwidth]{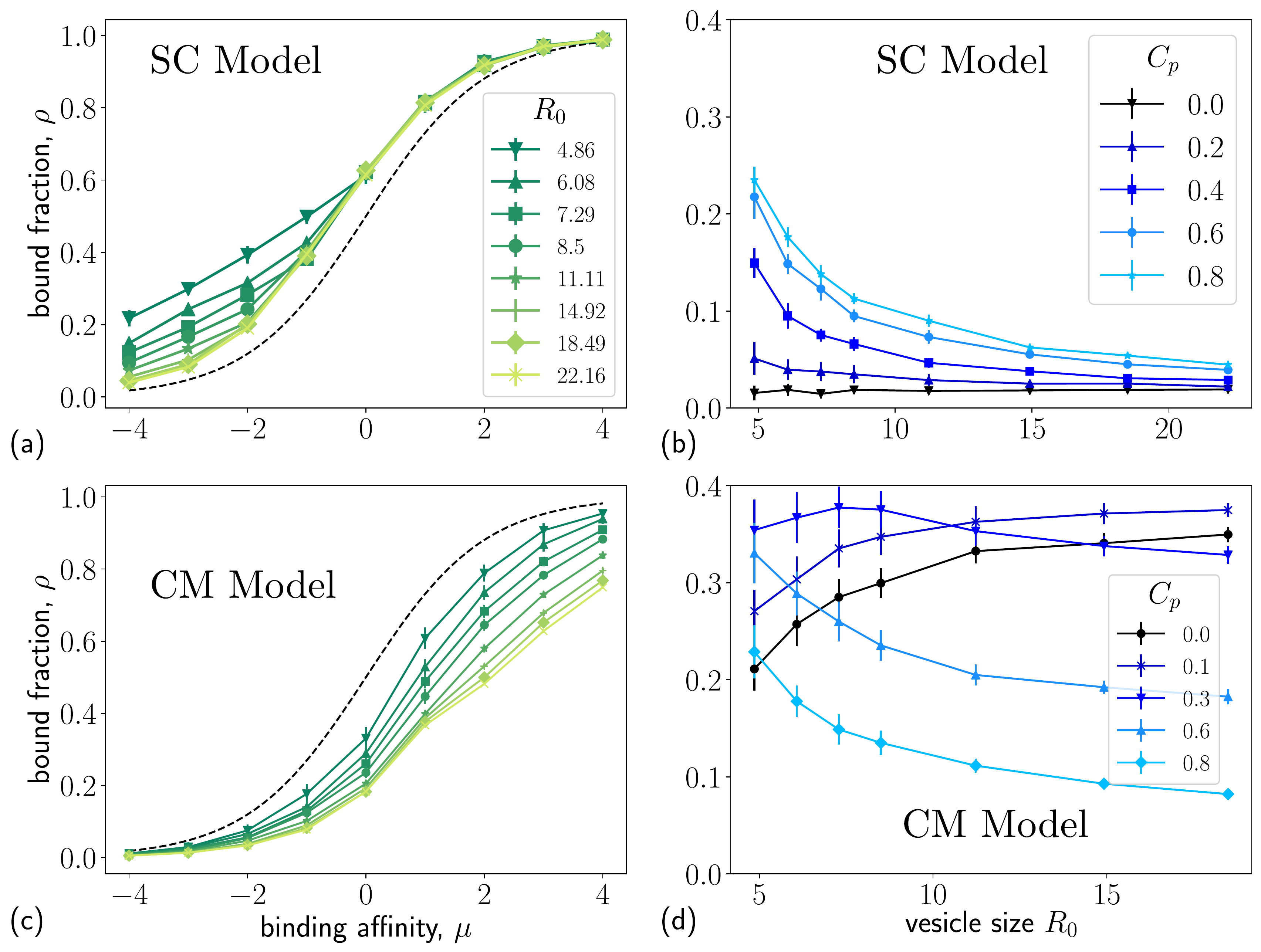}
  \caption{Analysis of protein binding on a deformable sphere modeled using the spontaneous curvature (SC) model and the curvature mismatch (CM) model.
    Adsorption isotherms for vesicles of various sizes at $C_p = 0.6$ for (a) SC model and (c) CM model.
%    At low binding affinity, the protein bound fraction depends on the vesicle size (curvature sensing regime),
%    whereas at high binding affinity, $\rho$ is independent of $\mu$ (curvature generation regime).
Curvature sensing curves for the SC model for different protein spontaneous curvatures (b) at $\mu = -4$ for the SC model and (d) at $\mu = 0$ for the CM model.}\label{fig:simulation}
\end{figure*}

One can ask the  question -- what is the size of the vesicle that shows maximum protein binding for proteins with fixed intrinsic curvature ($C_p$) at a given concentration ($\mu$)?
We see that for the SC model, the protein bound fraction is maximum as $H \to \infty$ or in other words, for the smallest vesicle.
On the other hand, the CM model predicts that the maximum binding is when the vesicle radius is $C_p^{-1}$, \textit{ie.} $H = C_p / 2$.
Essentially, the observed difference between the two models can be attributed to the fact that, in the CM model, a bound protein, in addition to inducing curvature, also adds to the membrane stiffness.

The analysis presented above is restricted to vesicles of fixed size and shape or in other words, the shape of the vesicle is  assumed  to  not  change on  protein binding.
The curvature generation by proteins is completely neglected because analytical minimization of the free energy is complicated when we allow for both the local mean curvature ($H$) and the protein bound state ($\phi$) to vary.
Therefore, in the subsequent section, we use computer simulations to perform this minimization where both curvature sensing and curvature generation by proteins are accounted for.

\section{Curvature sensing and generation}
\label{sec:simulations}
We employed dynamic triangulation monte carlo (DTMC) simulations with protein binding,  in the grand canonical ensemble,  as described in Ref.~\cite{Sachin2019}.
At any instant of the simulation, vesicles  are represented by a triangulated surface, whereas proteins are represented by an occupation number defined at the vertices on the triangulated surface. The simulations are carried out using   both the SC and CM models.

The adsorption isotherm obtained using the SC model is shown in Fig.~\ref{fig:simulation}a.
At low $\mu$, we see that the protein bound fraction depends on the vesicle size at a fixed binding affinity.
This is referred to as the curvature sensing regime.
At high values of $\mu$, the protein binding fraction is independent of vesicle size.
This is the curvature generation regime~\cite{Sachin2019}.
The adsorption curve for the CM model with $\bar{\kappa} = 10$ is shown in Fig.~\ref{fig:simulation}c.
Although the adsorption isotherm appears to be Langmuir-like at small binding affinities, it significantly deviates from the Langmuir behavior at higher $\mu$ values.
For the SC model, curvature sensing happens at low binding affinity, whereas for the CM model, curvature sensing is more at higher binding affinities.

Curvature sensing is quantitatively measured using the equilibrium bound fraction of proteins for different vesicle sizes at the same binding affinity.
Curvature sensing curve, from  the SC model,  monotonically increases with decreasing radius (see Fig.~\ref{fig:simulation}b), which is
 qualitatively similar to the predictions with non-deforming vesicles.
At  $C_p=0$, the bound fraction is independent of the vesicle radius, \textit{ie}. there is no curvature sensing.
For non-zero $C_p$, the bound fraction is maximum in the limit of zero radius.
On the other hand,  in the curvature sensing curve for the CM model,  shown in Fig.~\ref{fig:simulation}d,
proteins with $C_p = 0.0$  is also coupled to the membrane curvature and  senses it  with more binding  on larger vesicles.
The simulation results show that, for $C_p\ne0.0$, protein binding is maximum at a finite non-zero vesicle radius.
When $C_p = 0.3$, there is a clear  maximum at $R_0\approx 7.0$.
We expect that such a maxima exists for other non-zero $C_p$ values, however they fall outside the range of vesicle radius studied in our simulations.
Here again, the curvature sensing curves are qualitatively similar to the curves obtained using the analytical model.

\section{Concluding remarks}
\label{sec:conclusions}
The main differences in results from the two models can be summarized as follows:
\begin{itemize}
  \item SC model has a monotonic curvature sensing behavior, while the CM model has a non-monotonic sensing curve.
  \item The SC model has a curvature sensing regime at low $\mu$ and a curvature generation regime at high $\mu$, whereas the CM model shows curvature sensing for all $\mu$ values explored here.
  \item the $C_p = 0$ case does not sense curvature in the SC model, while in the CM model, proteins show sensing behavior at all $C_p$ values.
\end{itemize}
 The curvature sensing behavior is observed when the membrane is stiff.
  In the case of deformable vesicles, the binding of proteins leads to a softening of the membrane in both SC and CM models.
  In the CM model, there is also a term that rescales the effective bending modulus of the membrane with protein binding (see Eq.~(\ref{eq:cm_hamiltonian})).
  Thus, the softening is significantly higher for SC model than the CM model.
  Consequently, for the same $\mu$, protein binding is always higher for the SC model than the CM model.
  At high $\mu$ in the SC model, the membrane is soft enough to conform to any protein curvature and hence we do not see curvature sensitivity.
  On the other hand, for the CM model, the membrane does not become soft enough to allow curvature  generation even at high $\mu$.

In the SC model, the coupling between protein density and membrane  elasticity is only through the parameter   $C_p$ (refer Eq.~(\ref{eq:sc_hamiltonian})), which  serves as the  source for  curvature generation and sensing.
At low $C_p$ values, curvature sensing and generation is weak due to the weak coupling.
Such a model is probably adequate for peripheral proteins that generate curvature through the hydrophobic insertion mechanism, where the strength of the coupling and the curvature generated are directly related.
In the CM model, on the other hand, the parameter $C_p$  has two contributions to the membrane elasticity.
As in SC model, here too $C_p$ serves as  the coupling strength between the membrane curvature and protein density.
In addition,  it  couples  the membrane stiffness  to the  protein concentration through the $\bar{\kappa}$ term in Eq.~(\ref{eq:cm_hamiltonian}).
Experimentally, such a scenario arises when the dominant interaction with  the membrane is coming  from a laterally  extended region on the protein, say  a  region of charge leading to  electrostatic binding.
A recent finite element analysis of curvature generation on a 3D linear elastic membrane has proposed that electrostatic interaction is essential for curvature generation by BAR domains~\cite{Mahata2017}.
Thus, the CM model may be more appropriate to model peripheral proteins that generate curvature using the scaffolding and other mechanisms such as oligomerization or steric repulsion or in modeling transmembrane proteins.

%%Use section* for acknowledgements
\section*{Acknowledgement}
TVSK thanks IIT Palakkad for hospitality and computational resources.
The authors thank Department of Biotechnology, Ministry of Science and Technology, Govt.\ of India for the financial support through grant no. BT/PR8025/BRB/10/1023/2013.

%\balance
%%use \balance somewhere in the left column of the last page to balance the two columns in the end page

%%References section

\end{document}